\documentclass[superscriptaddress,showpacs,preprintnumbers,amsmath,amssymb,twocolumn]{revtex4}

\usepackage{graphicx,dcolumn,bm,mdwlist}

\begin{document}

\title{Broken axisymmetry phase of a spin-1 ferromagnetic Bose-Einstein
condensate}

\author{Keiji Murata}
\affiliation{Department of Physics, Tokyo Institute of Technology, Tokyo
152-8551, Japan}
\author{Hiroki Saito}
\affiliation{Department of Applied Physics and Chemistry, The University
of Electro-Communications, Tokyo 182-8585, Japan}
\author{Masahito Ueda}
\affiliation{Department of Physics, Tokyo Institute of Technology, Tokyo
152-8551, Japan}
\affiliation{ERATO, Japan Science and Technology Corporation (JST),
Saitama 332-0012, Japan}

\date{\today}

\begin{abstract}
A spin-1 ferromagnetic Bose-Einstein condensate subject to a certain
magnetic field exhibits a broken-axisymmetry phase in which the
magnetization tilts against the applied magnetic field due to the
competition between ferromagnetism and linear and quadratic Zeeman
effects.
The Bogoliubov analysis shows that in this phase two Goldstone modes
associated with U(1) and SO(2) symmetry breakings exist, in which phonons
and magnons are coupled to restore the two broken symmetries.
\end{abstract}


\maketitle

\section{introduction}

Bose-Einstein condensates (BECs) with spin degrees of freedom have
attracted growing attention since the first observation of the spin-1
$^{23}$Na BEC by the MIT group~\cite{D.M.Stamper-Kurn et al.,J.Stenger et
al.}.
In contrast to a magnetic trap, in which hyperfine-spin degrees of freedom
are frozen, an optical trap can confine atoms in all magnetic sublevels of
spin, allowing study of the magnetic properties of BECs.
A variety of experiments have been performed to date, on areas such as
spin domains~\cite{H.-J.Miesnr et al.}, interdomain
tunneling~\cite{tunneling}, and realization of a spin-2 $^{23}$Na
BEC~\cite{A.Gorllitz et al.}.
The spin-exchange dynamics of $^{87}$Rb BECs have been investigated
experimentally by Schmaljohann \textit{et al.}~\cite{H.Schmaljohann et
al.}, Chang \textit{et al.}~\cite{M.-S. Chang et al.}, and Kuwamoto
\textit{et al.}~\cite{T.Kuwamoto et al.}.
	
Theoretical investigations of the spinor BEC have also been carried out
extensively.
Mean field theory (MFT) for a spin-1 BEC was formulated by Ho~\cite{T-L
Ho} and Ohmi and Machida~\cite{T.Ohmi&K.Machida}.
The MFT of a spin-2 BEC was developed by Ciobanu \textit{et
	al.}~\cite{Ciobanu et al.} and Ueda and Koashi~\cite{Ueda&Koashi}.
Law \textit{et al.}~\cite{Law et al.} developed a many-body theory of
spin-1 antiferromagnetic BEC.
Koashi and Ueda~\cite{M.Koashi&M.Ueda} and Ho and Yip~\cite{Ho&Yip}
extended it to including the linear Zeeman effect and found that an
antiferromagnetic BEC realize a fragmented BEC for a weak magnetic field.
The Bogoliubov analysis was carried out by Huang and Gou~\cite{Huang&Gou}
and by Ueda~\cite{M.Ueda} in the presence of the linear Zeeman effect.
Their results agree with those obtained using a diagrammatic method by
Sz\'{e}pfalusy and Szirmai~\cite{Peter et al.}.
In these studies, the Zeeman effects are restricted to those up to the
linear order in the magnetic field.
A unique feature of trapped atomic systems is that linear and quadratic
Zeeman effects can be manipulated independently due to spin conservation.
If we take the quadratic Zeeman term into account, the ground-state 
phase diagram becomes much richer as shown in Ref.~\cite{J.Stenger et
al.}.
In particular, under a certain range of linear and quadratic Zeeman
effects, there is a special phase in which the magnetization tilts against
the applied magnetic field.
The investigation of some of the unique features of this phase is the
primary purpose of our study.

When a weak magnetic field is applied along the quantization axis, the $m
= 1$ or $-1$ state is favorable for a spin-1 ${}^{87}$Rb atom due to the
linear Zeeman effect and the ferromagnetic interaction, where $m$ refers
to the magnetic sublevel.
On the other hand, the quadratic Zeeman effect raises the energy of the
$m = \pm 1$ states relative to that of the $m = 0$ state.
As a consequence, if the quadratic Zeeman effect is sufficiently large,
the spin vector of the ferromagnetic ground state not only shrinks but
also tilts against the direction of the magnetic field.
Therefore, even if the Hamiltonian is axisymmetric with respect to the 
direction of the magnetic field, the ground state spontaneously breaks 
the axisymmetry.
This phase, which we shall refer to as a broken-axisymmetry phase, was
predicted in Ref.~\cite{J.Stenger et al.}, but little attention has been
paid to it from the viewpoint of axisymmetry breaking.

In the present study, we investigate the Goldstone modes of this phase by
studying its excitation spectrum.
The BEC with ferromagnetic interactions has three phases: ferromagnetic,
polar, and broken-axisymmetry phases.
In the ferromagnetic and polar phases, only the U(1) (global phase)
symmetry is broken, and the Goldstone mode corresponds to a phonon.
In the broken-axisymmetry phase, the SO(2) symmetry (axisymmetry) of the
spin vector is broken in addition to the U(1) symmetry.
Because of the simultaneous breaking of these two continuous symmetries,
the associated Goldstone modes are expected to involve both phonons and
magnons.

This paper is organized as follows.
Section~\ref{section2} reviews the mean-field ground state of a spin-1 BEC
to make this paper self-contained.
Section~\ref{section3} uses the Bogoliubov theory to derive one gapful
mode and two gapless Goldstone modes.
Section~\ref{sectiond} explores the implications of the present study for
other related studies, and Sec.~\ref{section4} concludes this paper.
Appendix~\ref{app0} derives analytic expressions for the
broken-axisymmetry phase.
Appendix~\ref{app} discusses excitations in the ferromagnetic and polar
phases for comparison with those in the broken-axisymmetry phase.

\section{Ground state with broken axisymmetry}
\label{section2}

\subsection{Formulation of the problem}
We consider a uniform system of $N$ identical bosons with hyperfine spin
1 in which an external magnetic field is applied in the $z$ direction.
The Hamiltonian of the system is written as the sum of one-body part
$\hat{\mathcal{H}}_\mathrm{I}$ and two-body interaction part
$\hat{\mathcal{H}}_\mathrm{II}$.
The one-body part is given by 
\begin{align}
    \hat{\mathcal{H}}_\mathrm{I} 
    = 
    \sum_{m=-1}^1
    \int \mathrm{d} \mathbf{r} 
    \hat{\Psi}_m^\dag
    \left( 
        -\frac{\hbar^2}{2M} \nabla^2 - p m + q m^2
    \right)
    \hat{\Psi}_m,
\label{hamiltonian1}
\end{align}
where subscripts $m=+1,0,-1$ denote the magnetic quantum numbers along the
$z$ axis, $M$ is the mass of the atom, and $p$ and $q$ are the linear and
quadratic Zeeman coefficients, respectively.
In the case of spin-1 ${}^{23}$Na and ${}^{87}$Rb atoms, $q$ is positive.
The two-body part, which is described by a contact-type $s$-wave
interaction at ultralow temperature, takes the form
\begin{align}
	\hat{\mathcal{H}}_\mathrm{II} 
    = 
    \frac{1}{2}
       \sum_{F=0,2}
       g_F 
    \int \mathrm{d} \mathbf{r} \hat{\Psi}_{n'}^\dag \hat{\Psi}_{m'}^\dag
	\langle m' ; n' |\mathcal{P}_F | m ; n \rangle
    \hat{\Psi}_m \hat{\Psi}_n, 
\label{hamiltonian2}
\end{align} 
where $g_F = 4 \pi \hbar^2 a_F/M$ with
$a_0$ and $a_2$ being the $s$-wave scattering lengths in the singlet and
quintuplet channels, respectively, and $\mathcal{P}_F$ projects a two-body
state into that with total spin $F$.
The absence of the projection onto the $F=1$ channel is due to the Bose
statistics.

Because the system is uniform, it is convenient to expand the field
operators in terms of plane waves as
$
    \hat{\Psi}_m = \Omega^{-1/2} \sum_{\mathbf{q}}
    e^{i \mathbf{q} \cdot \mathbf{r}}
    \hat{a}_{\mathbf{q}, m}
$,
where $\Omega$ is the volume of the system and $\hat{a}_{\mathbf{q}, m}$
represents the annihilation operator of a boson with wavenumber
$\mathbf{q}$ and magnetic sublevel $m$.
Equations (\ref{hamiltonian1}) and (\ref{hamiltonian2}) are then rewritten
as
\begin{align}
	\hat{\mathcal{H}}_\mathrm{I}= &
    \sum_{\mathbf{k},m}
    \left(
        \epsilon_{\mathbf{k}} -p m + q m^2 
    \right)
    \hat{a}_{\mathbf{k},m}^\dag \hat{a}_{\mathbf{k},m},
    \label{hamiltonian-3}\\
    \hat{\mathcal{H}}_\mathrm{II}= &
    \frac{c_0}{2 \Omega}
    \sum_{\mathbf{k}}
        : \hat{\rho}_{\mathbf{k}}^\dag \hat{\rho}_{\mathbf{k}} :
    + \frac{c_1}{2 \Omega}
    \sum_{\mathbf{k}}
        : \hat{\mathbf{f}}_{\mathbf{k}}^\dag \cdot \hat{\mathbf{f}}_{\mathbf{k}} :,
	\label{hamiltonian3}
\end{align} 
where $\epsilon_{\mathbf{k}} = \hbar^2 k^2/(2M)$, $c_0 = (g_0 + 2 g_2)/3$,
$c_1 = (g_2 - g_0)/3$,
$
	\hat{\rho}_{\mathbf{k}}
    = \sum_{\mathbf{q},m} \hat{a}_{\mathbf{q}, m}^\dag
                        \hat{a}_{\mathbf{q}+\mathbf{k}, m}
$,
and 
$
    \hat{\mathbf{f}}_{\mathbf{k}}
    = \sum_{\mathbf{q},m,n} \mathbf{f}_{m,n} \hat{a}_{\mathbf{q}, m}^\dag
                        \hat{a}_{\mathbf{q}+\mathbf{k}, n}
$
with $\mathbf{f} = (f_x, f_y, f_z)$ being the spin-1 matrices in vector
notation.
The symbol $: :$ denotes the normal ordering of operators.
The spin-spin interaction is ferromagnetic if $c_1 < 0$ and
antiferromagnetic if $c_1 >0$.
It is known that the interaction between spin-1 $^{23}$Na atoms is
antiferromagnetic and that the interaction between spin-1 $^{87}$Rb atoms
is ferromagnetic~\cite{Klausen et al.,E.G.M. van Kempen et al.}.

Assuming that a macroscopic number of atoms occupy the
$\mathbf{k}=\mathbf{0}$ state, we replace the relevant operators with
c-numbers.
The Hamiltonian for the BEC in the $\textbf{k}=\textbf{0}$ state is given
by
\begin{eqnarray}
	\hat{\mathcal{H}}_{\mathrm{BEC}}
&	= &
	\frac{c_0}{ 2\Omega } 
	:
		\left(
			\hat{a}_{\textbf{0},m}^\dag \hat{a}_{\textbf{0},m}
		\right)^2
	: \nonumber \\
& & +
	\sum_{m=-1}^{1} 
	\left(
		-pm +qm^2
	\right)
	\hat{a}_{\mathbf{0}, m}^\dag \hat{a}_{\mathbf{0}, m}
	-\frac{c_1}{2\Omega} \hat{s}^\dag \hat{s}, \nonumber \\
\label{Hbec}
\end{eqnarray}
where 
$
	\hat{s}=
	\left(
		\hat{a}_{\mathbf{0}, 0}^2 -2 \hat{a}_{\mathbf{0}, 1} \hat{a}_{\mathbf{0}, -1}
	\right)/\sqrt{3}
$ 
is an annihilation operator for a singlet pair.
In the MFT, we replace the operator $\hat{a}_{\textbf{0},m}$ with a
c-number $\zeta_m \sqrt{N_0}$.
Here, $N_0$ is the number of condensed atoms and the order parameters
$\zeta_m$'s are complex variational parameters that are determined so as
to minimize the energy functional under the constraint of normalization
$\sum_{m} |\zeta_m|^2 = 1$.
For this purpose, we introduce a Lagrange multiplier $\mu$ and minimize 
$
	\left \langle \mathcal{H} \right \rangle 
	- 
	\mu N_0 \sum_m |\zeta_m|^2 
$ 
with respect to $\zeta_m$.
In the following, we denote the set of the order parameters as
$\bm{\zeta}= {}^{T} (\zeta_1, \zeta_0, \zeta_{-1})$, where the superscript
$T$ stands for transpose.

\subsection{Ground states}

\begin{figure}[t]
	\begin{center}
	\includegraphics[height=12\baselineskip]{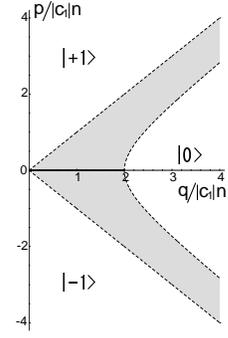}
	\end{center}
	\caption{
Ground-state phase diagram for a spin-1 ferromagnetic BEC.
The dashed curves indicate second-order phase boundaries.
In the figure, $\left|+1 \right\rangle$ and $\left|-1\right\rangle$
represent the ferromagnetic phase, and $\left|0\right\rangle$ represents
the polar phase.
The shaded region is the broken-axisymmetry phase, in which the
magnetization tilts against the $z$ axis.
}
	\label{fig:eps.eps}
\end{figure}

\begin{figure}[t]
	\begin{center}
	\includegraphics[height=9\baselineskip]{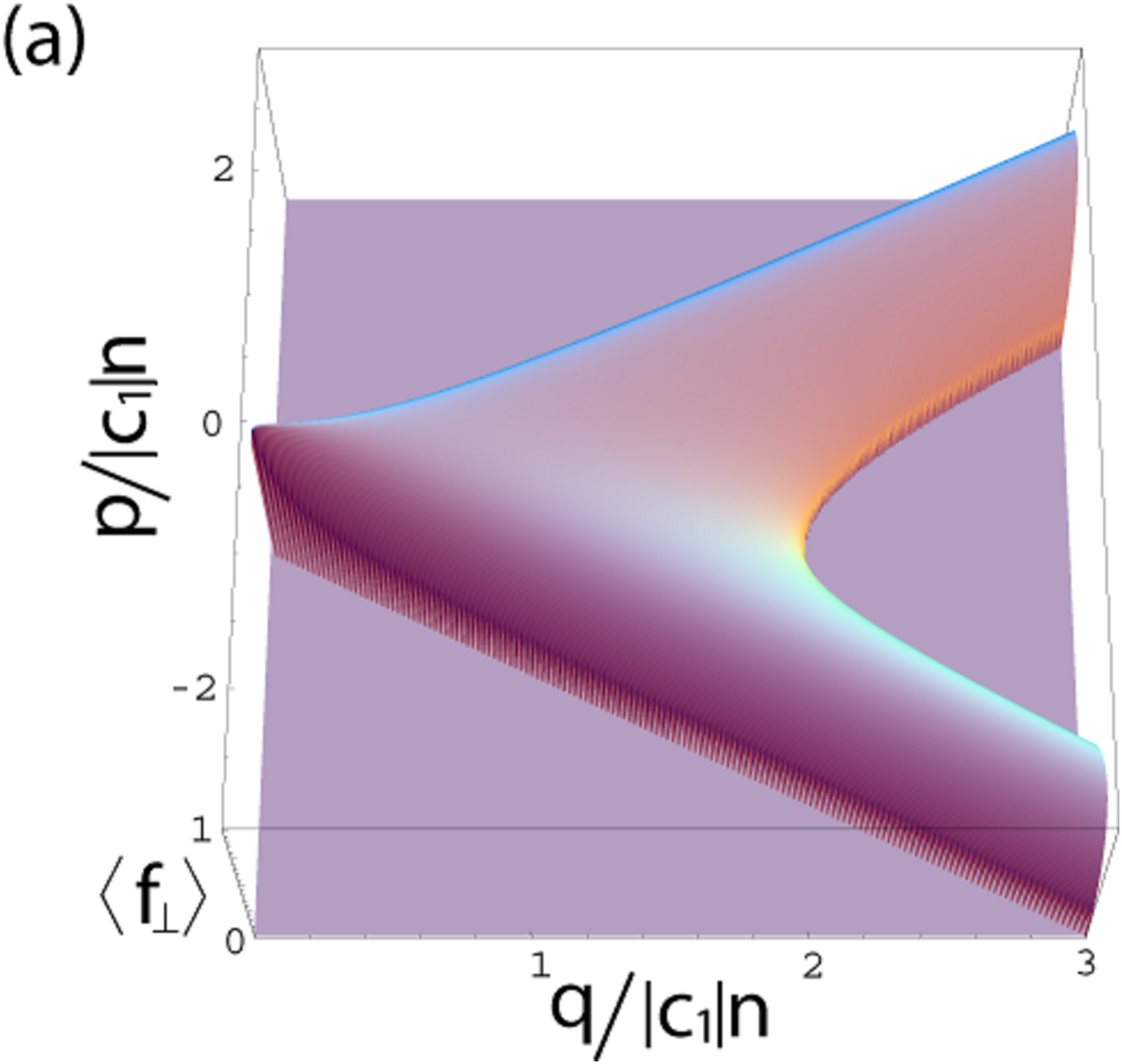}
	\includegraphics[height=9\baselineskip]{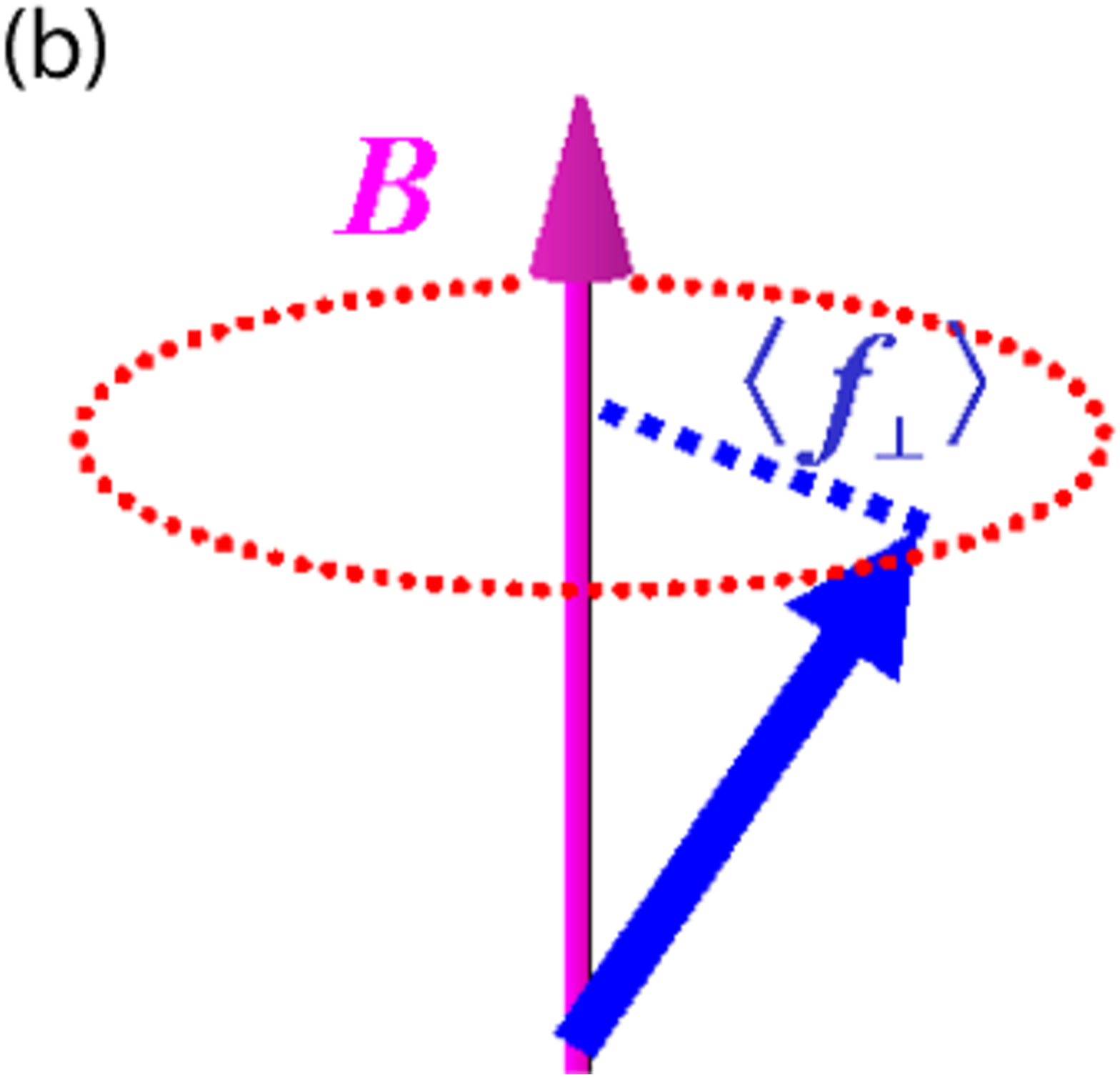}
	\end{center}
	\caption{ (Color online)
(a) Transverse magnetization, i.e., magnetization perpendicular to the
direction of the applied magnetic field $\langle f_\perp \rangle \equiv
\langle F_\perp \rangle / N_0$ as a function of linear and quadratic
Zeeman coefficients.
The transverse magnetization is nonzero only in the broken-axisymmetry
phase.
(b) Schematic illustration of the spin vector in the broken-axisymmetry
phase.
		}
	\label{f_perpendicular}
\end{figure}

The ground-state phase diagram for a spin-1 ferromagnetic BEC is shown in
Fig.~\ref{fig:eps.eps}~\cite{J.Stenger et al.}.
The phases are classified as follows:
\begin{enumerate}
\item Ferromagnetic phase ( $|+1\rangle $ and $|-1\rangle $ in
Fig.~\ref{fig:eps.eps}).
The order parameter is given for $p > 0$ by $\bm{\zeta}_{\mathrm{F}}=
{}^{T} (e^{i \chi_1}, 0, 0) $ and for $p < 0$ by
$\bm{\zeta}_{\mathrm{F}}={}^{T}(0, 0, e^{i \chi_{-1}})$, where $\chi_m$
denotes an arbitrary phase of $\zeta_m$, i.e., $\zeta_m = |\zeta_m| e^{i
\chi_m}$.
\item Polar phase ($|0\rangle $ in Fig.~\ref{fig:eps.eps}).
The order parameter is given by $\bm{\zeta}_{\mathrm{P}}= {}^{T} (0, e^{i
\chi_0}, 0)$.
\item Broken-axisymmetry phase (shaded region in Fig.~\ref{fig:eps.eps}).
The order parameter is given by (see Appendix~\ref{app0} for derivation)
\begin{align}
    \begin{cases}
        \zeta_{\pm 1} 
        = \left( q \pm p \right)
        \sqrt{
            \displaystyle
            \frac{p^2 + 2 \left| c_1 \right| n q - q^2}
                {8 \left| c_1 \right| n q^3}
            }
            e^{ i \chi_{\pm 1}},\\
        \zeta_0
        =\sqrt{
            \displaystyle
            \frac{
                \left(
                    q^2 - p^2
                \right)
                \left(
                    p^2 + 2 \left| c_1 \right| n q + q^2
                \right)
                }
                {4 \left| c_1 \right| n q^3}
            }
            e^{i (\chi_1 + \chi_{-1})/2}.
    \end{cases}
    \label{mixed}
\end{align}
\end{enumerate}
In the broken-axisymmetry phase, the transverse magnetization, which is
perpendicular to the external magnetic field,
\begin{eqnarray}
\label{fperp}
    \left \langle F_\perp \right \rangle
& \equiv & \sqrt{\left\langle F_x \right\rangle^2 + \left\langle F_y
\right\rangle^2}
\nonumber \\
    & = &
    N_0\frac{
            \sqrt{q^2 - p^2} 
            \sqrt{
                    \left(
                        p^2 + 2 \left| c_1 \right| n q
                    \right)^2
                    -q^4
                }
        }
        {
            2 \left| c_1 \right| n q^2
        },
\end{eqnarray}
is nonzero as shown in Fig.~\ref{f_perpendicular}.
(If we choose $\zeta_0$ to be real and positive, we have
$
    \left \langle F_x \right \rangle
    =
    \left \langle F_\perp \right \rangle \cos \phi
$ and
$
    \left \langle F_y \right \rangle
    =
    \left \langle F_\perp \right \rangle \sin \phi
$,
where $\phi \equiv \chi_1 = -\chi_{-1}$.)
The longitudinal magnetization, which is parallel to the external magnetic
field, is given by
\begin{equation}
\left\langle F_z \right\rangle = N_0 \frac{p \left( p^2 + 2 \left| c_1
\right| n q - q^2 \right)}{2 \left| c_1 \right| n q^2}.
\end{equation}
The total magnetization is therefore given by
\begin{equation} \label{fmag}
\left| \left\langle \bm{F} \right\rangle \right| \equiv \sqrt{\left\langle
F_\perp \right\rangle^2 + \left\langle F_z \right\rangle^2} = N_0
\frac{\sqrt{4 c_1^2 n^2 q^2 - \left( p^2 - q^2 \right)^2}}{2
\left| c_1 \right| n q}.
\end{equation}
The magnetization thus tilts against the applied magnetic field with
the polar angle
\begin{align} \label{pangle}
	\vartheta = \arctan 
	\left(
		\frac{ \sqrt{q^2 - p^2} \sqrt{p^2 + 2 \left| c_1 \right| n q + q^2} }
			{ p \sqrt{p^2 + 2 \left| c_1 \right| n q - q^2} }
	\right).
\end{align}

We note that this ground state breaks the axisymmetry around the $z$ axis
despite the fact that the Hamiltonian including the external magnetic
field is axisymmetric.
Thus, the ground state features spontaneous breaking of axisymmetry or
spontaneous breaking of the SO(2) symmetry.
Such an axisymmetry breaking is due to the competition between the linear
and quadratic Zeeman effects and the ferromagnetic interaction.
The quadratic Zeeman effect decreases the $z$ component of the spin
vector.
However, a decrease in the length of the spin vector costs the
ferromagnetic interaction energy.
To reconcile the quadratic Zeeman effect with the ferromagnetic
interaction, the spin vector tilts against the $z$ axis.
In fact, $\vartheta$ in Eq.~(\ref{pangle}) is a monotonically decreasing
function of $p$ and a monotonically increasing function of $q$, and the
length of the spin vector (\ref{fmag}) attains the maximum value of $N_0$
for $|c_1| n \rightarrow \infty$.

\section{Bogoliubov excitations and Goldstone modes}
\label{section3}

According to the Goldstone theorem~\cite{Goldstone}, there exists a
gapless excitation mode when a continuous symmetry is spontaneously
broken.
In the broken-axisymmetry phase, we have shown that the relevant
continuous symmetry is the SO(2) axisymmetry.
Since we have treated the system in the MFT in which the global phase of
the wave function is assumed to be arbitrarily chosen, the U(1) symmetry is
also broken.
Thus, the two continuous symmetries are simultaneously broken in this
phase.
In this section, we examine the corresponding Goldstone modes using the
Bogoliubov theory.
	
\subsection{Basic theory}

We first formulate a number-conserving Bogoliubov
theory~\cite{Bogoliubov} for a BEC with spin degrees of
freedom~\cite{M.Ueda}.
The advantage of this number-conserving formulation is that we do not need
to introduce the chemical potential as a Lagrange multiplier to adjust the
particle number to a prescribed value.
In this formulation, we replace $\hat{a}_{\mathbf{0}, m}$ with $\zeta_m
\left( N - \sum_{\textbf{k}\neq\textbf{0},m} \hat{a}_{\mathbf{k}, m}^\dag
\hat{a}_{\mathbf{k}, m} \right)^{1/2}$ in Eqs.~(\ref{hamiltonian-3}) and
(\ref{hamiltonian3}) and keep terms up to those of second order in
$\hat{a}_{\mathbf{k}\neq \textbf{0}, m}$ and
$\hat{a}_{\mathbf{k}\neq\textbf{0}, m}^\dag$.
We then obtain an effective Bogoliubov Hamiltonian as 
\begin{widetext}
\begin{align}
			\hat{\mathcal{H}}_{\mathrm{eff}}
			= &
			\sum_{\textbf{k} \neq \textbf{0} }
			\sum_{m=-1}^{1}
			\left(
					\epsilon_{\textbf{k}}
					- p m
					+ q m^2
					+ p \langle f_z \rangle
					- q \langle f_z^2 \rangle
					- c_1 n 
					+ c_1 n | \zeta_0^2 - 2 \zeta_1 \zeta_{-1} |^2
			\right)
				\hat{a}_{ \textbf{k} , m }^\dag \hat{a}_{ \textbf{k} , m }
			\nonumber \\ &
			+ 
			c_1 n \langle \bm{f} \rangle \cdot
			\sum_{ \textbf{k} \neq \textbf{0} }\sum_{m,n} 
			\bm{f}_{m,n}
			\hat{a}_{ \textbf{k} , m }^\dag
			\hat{a}_{ \textbf{k} , n }
			+
			\frac{c_0 n}{2}
			\sum_{ \textbf{k} \neq \textbf{0} } 
			\left(
					2 \hat{\mathcal{D}}_{\textbf{k}}^\dag \hat{\mathcal{D}}_{\textbf{k}}
					+
					\hat{\mathcal{D}}_{\textbf{k}} \hat{\mathcal{D}}_{-\textbf{k}}
					+
					\hat{\mathcal{D}}_{\textbf{k}}^\dag \hat{\mathcal{D}}_{-\textbf{k}}^\dag
			\right)
			\nonumber \\ &
			+
			\frac{c_1 n}{2} 
			\sum_{\textbf{k} \neq \textbf{0} }
				\left(
						2 \hat{\bm{\mathcal{F}}}_{\textbf{k}}^\dag \cdot 
						\hat{\bm{\mathcal{F}}}_{\textbf{k}}
						+
						\hat{\bm{\mathcal{F}}}_{\textbf{k}} \cdot 
						\hat{\bm{\mathcal{F}}}_{-\textbf{k}}
						+
						\hat{\bm{\mathcal{F}}}_{\textbf{k}}^\dag \cdot
						\hat{\bm{\mathcal{F}}}_{-\textbf{k}}^\dag
				\right)
			+E_0,
			\label{original Hamiltonian}
\end{align}  
\end{widetext}	
where	$\hat{\mathcal{D}}_{\textbf{k}} \equiv \sum_{m} \zeta_{m}^*
\hat{a}_{\textbf{k}, m}$, $\hat{\bm{\mathcal{F}}}_{\textbf{k}} \equiv
\sum_{m,n} \bm{f}_{m,n} \zeta_{m}^* \hat{a}_{\textbf{k}, n}$, and $E_0$
represents a constant term.
	
In general, for spin-$f$ atoms, we can express quasiparticle operators
$\hat{b}_{\mathbf{k}, \sigma}$'s as linear combinations of the
annihilation and creation operators of the original particles:
\begin{align}
    	\hat{\mathbf{B}}_{\textbf{k}} 
    	= \mathrm{U}(k) \hat{\mathbf{A}}_{\textbf{k}} 
    	+ \mathrm{V}(k) \hat{\mathbf{A}}_{-\textbf{k}}^*.
    	\label{defofqp}
\end{align}
Here $\mathrm{U}(k)$ and $\mathrm{V}(k)$ are $(2f+1) \times (2f+1)$ real
matrices and the bold letters represent sets of operators
\begin{align*}
    	\hat{\mathbf{B}}_{\textbf{k}}&= \,^{T} \! \left(
        	\hat{b}_{\mathbf{k},\sigma_1} \, , \,
        	\hat{b}_{\mathbf{k},\sigma_2} \, , \,
        	\cdots \, , \,
        	\hat{b}_{\mathbf{k},\sigma_{2f+1}} 
    	\right),\\
   		\hat{\mathbf{A}}_{\textbf{k}}&= \,^{T} \! \left(
        	\hat{a}_{\mathbf{k},f} \, ,\, 
        	\hat{a}_{\mathbf{k},f-1} \, , \,
        	\cdots \, , \,
        	\hat{a}_{\mathbf{k},-f} 
    	\right),\\ 
    	\hat{\mathbf{A}}_{\textbf{k}}^*&= \,^{T} \! \left(
        	\hat{a}_{\mathbf{k},f}^\dag \, , \,
        	\hat{a}_{\mathbf{k},f-1}^\dag \, , \,
        	\cdots \, , \,
        	\hat{a}_{\mathbf{k},-f}^\dag 
    	\right),
\end{align*}
where $\sigma_j$ is the label for each Bogoliubov mode.
The quasiparticle operators (\ref{defofqp}) should satisfy the Bose
commutation relations,
\begin{align}
    	    \left[
    	        \hat{b}_{\mathbf{k}, \sigma}, \hat{b}_{\mathbf{k}', \sigma '}
    	    \right]
    	    =0 
    	    , \quad 
    	    \left[
    	        \hat{b}_{\mathbf{k}, \sigma},\hat{b}_{\mathbf{k}', \sigma '}^\dag
    	    \right]
    	    =
    	    \delta_{\mathbf{k},\mathbf{k}'}\delta_{\sigma , \sigma '},
		\label{com_rel}
\end{align}	
which lead to
\begin{eqnarray}
		\sum_i 
		\left[
			U_{\sigma , i}(k)\, ^T U_{i , \sigma'}(k)
			-
			V_{\sigma , i}(k)\, ^T V_{i , \sigma'}(k)
		\right] & = & \delta_{\sigma, \sigma'} ,
\nonumber \\
\label{comrelcomp1}
\\
		\sum_i 
		\left[
			U_{\sigma , i}(k)\, ^T V_{i , \sigma'}(k)
			-
			V_{\sigma , i}(k)\, ^T U_{i , \sigma'}(k)
		\right] & = & 0.
\label{comrelcomp2}
\end{eqnarray}
We can rewrite Eqs.~(\ref{comrelcomp1}) and (\ref{comrelcomp2}) in a
matrix form,
\begin{align}
	    \,^{T} \!
	    \left(
	        \mathrm{U} + \mathrm{V}
	    \right)
	    \left(
	        \mathrm{U} - \mathrm{V}
	    \right)
	    = \mathrm{I}.
	    \label{commutation relation mat}
\end{align} 
Thus, U and V are not independent of each other.
For later convenience, we write $\hat{\mathbf{B}}_{\textbf{k}}$ as 
\begin{align}
		\hat{\mathbf{B}}_{\textbf{k}} = 
		\frac{1}{2}
		\left[
		\left( 
			\mathrm{U} + \mathrm{V} 
		\right)
		\left( \hat{\mathbf{A}}_{\textbf{k}} 
		+ 
		\hat{\mathbf{A}}_{-\textbf{k}}^* \right)
		+
		\left( 
			\mathrm{U} - \mathrm{V} 
		\right)
		\left( \hat{\mathbf{A}}_{\textbf{k}} 
		- 
		\hat{\mathbf{A}}_{-\textbf{k}}^* \right)
		\right].
	\label{Re_Bogoliubov}
\end{align}

We seek the excitation spectrum $E_{\sigma}$ and operators
$\hat{b}_{\mathbf{k}, \sigma}$ such that the quasiparticles behave
independently, i.e.,
\begin{align}
    	\displaystyle \hat{\mathcal{H}}_{\mathrm{eff}}
    	    =\sum_{\mathbf{k} \neq 0}
    	    \sum_{\sigma = \left\{ \sigma_1,\sigma_2,\cdots,\sigma_{2f+1} \right\}}
    	        E_{\sigma} 
    	        \hat{b}_{\mathbf{k}, \sigma}^\dag \hat{b}_{\mathbf{k}, \sigma}
    	        +E_{\mathrm{vac}},
    	\label{free particle like}
\end{align}
where $\hat{\mathcal{H}}_{\mathrm{eff}}$ is given in Eq.~(\ref{original
Hamiltonian}), and $E_{\mathrm{vac}}$ is the energy of the vacuum state
for the quasiparticles.
From Eq.~(\ref{original Hamiltonian}), the Heisenberg equation of motion
takes the form
\begin{align}
    	i \hbar \frac{\mathrm{d}}{\mathrm{d}t} \hat{\mathbf{A}}_{\textbf{k}}
    	= \mathrm{M}(k) \hat{\mathbf{A}}_{\textbf{k}} 
    	+ \mathrm{N}(k) \hat{\mathbf{A}}_{-\textbf{k}}^*,
    	\label{differential of a}
\end{align}
where $\mathrm{M}(k)$ and $\mathrm{N}(k)$ are real and symmetric
$(2f+1) \times (2f+1)$ matrices.
Using the quasiparticle Hamiltonian (\ref{free particle like}) and the
commutation relations (\ref{com_rel}), on the other hand, we obtain 
\begin{align}
	    i \hbar \frac{\mathrm{d}}{\mathrm{d}t} 
	    \hat{\mathbf{B}}_{\textbf{k}}
	    = \mathrm{E}(k) \hat{\mathbf{B}}_{\textbf{k}},
	    \label{differential of b}
\end{align}
where $\mathrm{E}(k)$ is the diagonal $(2f+1) \times (2f+1)$ matrix, whose
diagonal elements correspond to the energies of the elementary excitations
$E_{\textbf{k},\sigma_j}$.
Then substituting Eq.~(\ref{differential of a}) into
Eq.~(\ref{differential of b}) and using Eq.~(\ref{commutation relation
	mat}), we obtain
\begin{align}
		    \left(
	    	    \mathrm{M} + \mathrm{N}
	    	\right)
	    	\left(
	    	    \mathrm{M} - \mathrm{N}
	    	\right)
	    	\,^{T} \!
	    	\left(
	    	    \mathrm{U} + \mathrm{V}
	    	\right)
	    	=
	    	\,^{T} \!
	    	\left(
	    	    \mathrm{U} + \mathrm{V}
	    	\right)
	    	\mathrm{E}^2 .
\end{align}
Since $\mathrm{E}^2$ is also a diagonal matrix, the Bogoliubov excitation
spectrum can be found as the eigenvalues of the matrix
\begin{align} \label{defG}
		\mathrm{G}
		\equiv
		\left(
	        \mathrm{M} + \mathrm{N}
	    \right)
	    \left(
	   	    \mathrm{M} - \mathrm{N}
	    \right).
\end{align}
We note that G, which is the product of two Hermitian matrices, is not, in
general, Hermitian.
The present approach has the advantage that we can reduce the dimension
of the eigenvalue equation from $2(2f+1)$ to $(2f+1)$ and therefore
the diagonalization is simplified.
That is, instead of the diagonalization of the $2(2f+1) \times 2(2f+1)$
matrix as
\begin{align}
		\begin{pmatrix}
			\mathrm{M} & \mathrm{N} \\
			-\mathrm{N} & -\mathrm{M}
		\end{pmatrix}
		\to
		\begin{pmatrix}
			\mathrm{E} & 0 \\
			0 & -\mathrm{E}
		\end{pmatrix},	
\end{align}
a $(2f+1) \times (2f+1)$ matrix G is to be diagonalized.

\subsection{Low-lying modes in the broken-axisymmetry phase for $\bm{k}
\rightarrow \bm{0}$}

Without loss of generality, we may assume $\zeta_m$ to be real and
positive.
The excitation spectra in the ferromagnetic and polar phases can be
derived analytically as shown in Appendix~\ref{app}.
The analytic solutions can also be obtained for the broken-axisymmetry
phase.
However, since they are very complicated, we here derive the excitation
spectrum for small $k$.

The effective Hamiltonian (\ref{original Hamiltonian}) gives the
coefficient matrices M and N of the Heisenberg equation of motion
(\ref{differential of a}).
Using the explicit form of $\zeta_m$ in Eq.~(\ref{mixed}), the matrix G
can be written in the form,
\begin{equation}
\mathrm{G} = \mathrm{G}_0 + 2(g_2 n \mathrm{G}_1 - c_1 n
\mathrm{G}'_1) \epsilon_{\mathbf{k}} +
I \epsilon_{\mathbf{k}}^2,
\end{equation}
where $I$ is the unit matrix and
\begin{align}
		\mathrm{G}_0 =& 
			\begin{pmatrix}
			\Theta_1 \zeta_{-1} \zeta_{0 } & 
			\Theta_1 \zeta_{ 1} \zeta_{-1} & 
			\Theta_1 \zeta_{ 1} \zeta_{0} \\
			\Theta_0 \zeta_{-1} \zeta_{0 } & 
			\Theta_0 \zeta_{ 1} \zeta_{-1} & 
			\Theta_0 \zeta_{ 1} \zeta_{0} \\
			\Theta_{-1} \zeta_{-1} \zeta_{0 } & 
			\Theta_{-1} \zeta_{ 1} \zeta_{-1} & 
			\Theta_{-1} \zeta_{ 1} \zeta_{0} 
			\end{pmatrix},
		\label{G0}
		\\
		\mathrm{G}_1 =& 
			\begin{pmatrix}
			\zeta_{ 1} ^2 & \zeta_{ 1} \zeta_{ 0} & \zeta_{ 1} \zeta_{-1} \\
			\zeta_{ 1} \zeta_{0 } & \zeta_{ 0} ^2 & \zeta_{-1} \zeta_{0} \\
			\zeta_{ 1} \zeta_{-1} & \zeta_{-1} \zeta_{ 0} & \zeta_{-1} ^2 
			\end{pmatrix},
		\label{G1}
		\\
		\mathrm{G}_1' =& 
			\begin{pmatrix}
			\zeta_{ 0} ^2 \zeta_{-1}/\zeta_{ 1} & -2 \zeta_{-1} \zeta_{ 0} & -2 \zeta_{ 1} \zeta_{-1} \\
			-2 \zeta_{-1} \zeta_{0 } & \zeta_{ 0} ^2 + 2 \zeta_{ 1} \zeta_{-1} & -2\zeta_{-1} \zeta_{0} \\
			-2\zeta_{ 1} \zeta_{-1} & -2\zeta_{-1} \zeta_{ 0} & \zeta_{ 0} ^2 \zeta_{ 1}/\zeta_{-1}
			\end{pmatrix},
		\label{eq:3matrices}
\end{align}
with
\begin{eqnarray}
		\Theta_m & = & (3m^2 - 2) |c_1| n
		\bigl[
			p^2 + 2 |c_1| n q + (-1)^{m} q^2 
\nonumber \\
& & - 2 m p q \bigr]
		\zeta_0 / (q \zeta_m).
\end{eqnarray}
We first consider the limit of $ \epsilon_{\mathbf{k}} \to 0$.
The eigenvalues for $\mathrm{G} = \mathrm{G}_0$ can be obtained easily:
one is
\begin{align}
		E_{\mathrm{gap}}^2
		=
		(3 p^2 - 2 c_1 n q -q^2)
		(p^2 - 2 c_1 n q + q^2)/q^2
\end{align}
and the other two are zero.
Thus the system has two gapless excitation modes, which, as we will show
later, arise from the U(1) and SO(2) symmetry breakings.
We label the gapful mode as $\alpha$, and the other two gapless modes as
$\beta$ and $\gamma$.
	
The eigenvectors of $\mathrm{G}_0$ are given by each row of the following
matrix:
\begin{align}
			\mathrm{U} + \mathrm{V} =
		\begin{pmatrix}
			\lambda \Theta_1 & \lambda \Theta_0 & \lambda \Theta_{-1} \\
			( \mu_{\mathrm{p} } + \mu_{\mathrm{m} } ) \zeta_1 &
			\mu_{\mathrm{p} } \zeta_0 &
			( \mu_{\mathrm{p} } - \mu_{\mathrm{m} } ) \zeta_{-1}\\
			( \nu_{\mathrm{p} } + \nu_{\mathrm{m} } ) \zeta_1 &
			\nu_{\mathrm{p} } \zeta_0 &
			( \nu_{\mathrm{p} } - \nu_{\mathrm{m} } ) \zeta_{-1}		
		\end{pmatrix},
		\label{uplusv}
\end{align}
where $\lambda$, $\mu_{\mathrm{p}}$, $\mu_{\mathrm{m}}$,
$\nu_{\mathrm{p}}$, and $\nu_{\mathrm{m}}$ are arbitrary parameters.
Note that the second and the third rows are the linear combinations of two
basis vectors $(\zeta_1, \zeta_0, \zeta_{-1})$ and $(\zeta_1, 0,
-\zeta_{-1})$, both of which are eigenvectors with zero eigenvalue.
It follows from Eq.~(\ref{commutation relation mat}) that the matrix 
$\mathrm{U} - \mathrm{V}$ is given as the transposed
inverse matrix of Eq.~(\ref{uplusv}),
\begin{widetext}
\begin{align}
			\mathrm{U} - \mathrm{V} = \frac{1}{A}
		\begin{pmatrix}
			\frac{ \zeta_{-1} \zeta_0 }{\lambda} &
			2 \frac{ \zeta_{1} \zeta_{-1} }{\lambda} &
			\frac{ \zeta_{1} \zeta_0 }{\lambda} \\
			\frac{ 
				-\zeta_{-1} \Theta_0 
					( \nu_{\mathrm{p} } - \nu_{\mathrm{m} } ) 
				+ \zeta_0 \Theta_{-1} \nu_{\mathrm{p} }
			}{J} &
			\frac{ 
				\zeta_{-1} \Theta_1 
					( \nu_{\mathrm{p} } - \nu_{\mathrm{m} } ) 
				- \zeta_1 \Theta_{-1}
					( \nu_{\mathrm{p} } + \nu_{\mathrm{m} } )
			}{J} &
			\frac{ 
				\zeta_{1} \Theta_0 
					( \nu_{\mathrm{p} } + \nu_{\mathrm{m} } ) 
				- \zeta_0 \Theta_{1} \nu_{\mathrm{p} }
			}{J} \\
			\frac{ 
				\zeta_{-1} \Theta_0 
					( \mu_{\mathrm{p} } - \mu_{\mathrm{m} } ) 
				- \zeta_0 \Theta_{-1} \mu_{\mathrm{p} }
			}{J} &
			\frac{ 
				- \zeta_{-1} \Theta_1 
					( \mu_{\mathrm{p} } - \mu_{\mathrm{m} } ) 
				+ \zeta_1 \Theta_{-1}
					( \mu_{\mathrm{p} } + \mu_{\mathrm{m} } )
			}{J} &
			\frac{ 
				- \zeta_{1} \Theta_0 
					( \mu_{\mathrm{p} } + \mu_{\mathrm{m} } ) 
				+ \zeta_0 \Theta_{1} \mu_{\mathrm{p} }
			}{J} 
		\end{pmatrix},
		\label{uminusv}
\end{align}
\end{widetext}
where $A \equiv 2\zeta_1 \zeta_{-1}\Theta_0 -\zeta_0 \zeta_{-1}\Theta_1 -
\zeta_0 \zeta_{1}\Theta_{-1}$ and $J \equiv \mu_{\mathrm{p}
}\nu_{\mathrm{m} }- \mu_{\mathrm{m} }\nu_{\mathrm{p} }$.

\subsection{Low-lying modes in the broken-axisymmetry phase for small
$\bm{k}$}

\begin{figure}[t]
		\begin{center}
		\includegraphics[width=19\baselineskip]{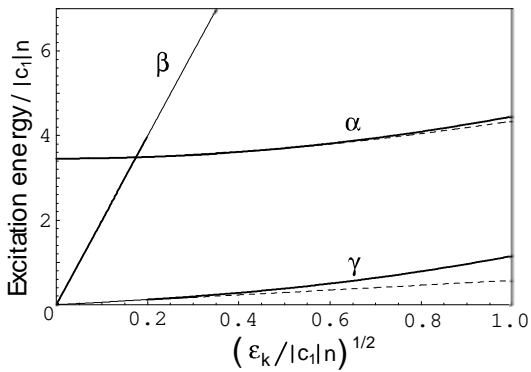}
		\end{center}
		\caption{
Excitation spectrum in the broken-axisymmetry phase.
The solid curves represent the exact solutions and the dashed ones are
approximate solutions given in Eq.~(\ref{three modes}).
The linear and quadratic Zeeman coefficients are chosen to be $p / |c_1| n
= 9 / 10$ and $q / |c_1| n = 11 / 10$.
The modes $\beta$ and $\gamma$ are the gapless modes associated with the
simultaneous breakings of U(1) and SO(2) symmetries.
The exact and approximate solutions for the $\beta$ mode cannot be
distinguished in this figure.
}
\label{fig:graph.eps}
\end{figure}

In the limit of small $\epsilon_{\mathbf{k}}$, the five parameters
$\lambda$, $\mu_{\mathrm{p}}$, $\mu_{\mathrm{m}}$, $\nu_{\mathrm{p}}$, and
$\nu_{\mathrm{m}}$ can be determined by substitution of
Eqs.~(\ref{uplusv}) and (\ref{uminusv}) into the definitions of the
Bogoliubov operators (\ref{Re_Bogoliubov}) and by comparison of the
quasi-particle Hamiltonian (\ref{free particle like}) with the effective
Hamiltonian (\ref{original Hamiltonian}).
However, we cannot perform this procedure using only the expression for
$\epsilon_{\textbf{k}} = 0$ because the two Goldstone modes diverge in the
limit of $\epsilon_{\textbf{k}} \to 0$.
Thus, it is necessary to find the $\epsilon_{\mathbf{k}}$ dependence of
the eigenenergies to find the properties of the low-lying excitations.
From Eqs.~(\ref{G0})-(\ref{eq:3matrices}), we obtain the eigenenergies up
to the order of $\epsilon_{\mathbf{k}}$ as
\begin{align}
		\begin{cases}
		E_{\alpha}^2 = E_{\mathrm{gap}}^2 +
		4 \left(
			\frac{ p^2 -c_1 n q }{q}
		\right)
		\epsilon_{\mathbf{k}} + 
		\mathrm{O} \left( \epsilon_{\mathbf{k}}^2 \right), 
		\\
		E_{\beta }^2 = \Lambda_{+} \epsilon_{\mathbf{k}} + 
		\mathrm{O} \left( \epsilon_{\mathbf{k}}^2 \right), 
		\\
		E_{\gamma}^2 = \Lambda_{-} \epsilon_{\mathbf{k}} + 
		\mathrm{O} \left( \epsilon_{\mathbf{k}}^2 \right),
	    \end{cases}
	    \label{three modes}
\end{align} 
where 
\[
		\Lambda_{\pm} = g_2 n + \frac{\eta}{2}
			\pm \frac{1}{2} 
			\sqrt{
				(2 g_2 n - \eta)^2 + \frac{8 g_2 n (q - \eta) \eta^2}
				{c_1 n (3 \eta - 2 q + 2 c_1 n)}
			}
\]
with $\eta = (q^2 -p^2)/q$.
In Fig.~\ref{fig:graph.eps}, we compare the $\epsilon_{\textbf{k}}$
dependences of the approximate eigenenergies (\ref{three modes}) (dashed
curves) with those of the numerically obtained exact energies (solid
curves).

It is important to note that the two gapless excitations $E_\beta$ and
$E_\gamma$ in Eq.~(\ref{three modes}) share the same leading-order
term $\epsilon_{\mathbf{k}}^{1/2}$.
Since the effective Hamiltonian $\hat{\mathcal{H}}_{\mathrm{eff}}$ in
Eq.~(\ref{original Hamiltonian}) contains only the terms that are
proportional to $\epsilon_{\mathbf{k}}$, this
$\epsilon_{\mathbf{k}}^{1/2}$ dependence must be canceled by the operators
$\hat{b}_{\mathbf{k},\sigma}$ so that Eq.~(\ref{free particle like})
reproduces Eq.~(\ref{original Hamiltonian}).
Therefore, we find that the normalization factors $\mu_{\mathrm{p}}$,
$\mu_{\mathrm{m}}$, $\nu_{\mathrm{p}}$, and $\nu_{\mathrm{m}}$ in
Eq.~(\ref{uplusv}), which determines $\hat{b}_{\mathbf{k},\sigma}$ through
Eq.~(\ref{defofqp}), must be proportional to either
$\epsilon_{\mathbf{k}}^{-1/4}$ or $\epsilon_{\mathbf{k}}^{1/4}$.
From numerical analysis, we find that they all have an
$\epsilon_{\mathbf{k}}^{-1/4}$ dependence as is the case in gapless
excitation modes in the other phases (see Appendix~\ref{app}).
It follows then from Eqs.~(\ref{uplusv}) and (\ref{uminusv}) that
$(\mathrm{U} + \mathrm{V})_{\sigma,m} \sim \mathrm{O}
(\epsilon_{\mathbf{k}}^{-1/4})$ and $(\mathrm{U} - \mathrm{V})_{\sigma,m}
\sim \mathrm{O} (\epsilon_{\mathbf{k}}^{1/4})$, and
\begin{align}
	(\mathrm{U} + \mathrm{V})_{\sigma,m} \gg (\mathrm{U} -
	\mathrm{V})_{\sigma,m} \;\;\;\;\; \mbox{($\sigma = \beta$ or $\gamma$)}
	\label{hermitian-like}
\end{align}
for $\epsilon_{\mathbf{k}} \to 0$.
Therefore, we can neglect the second term in the square bracket in
Eq.~(\ref{Re_Bogoliubov}), obtaining
\begin{align}
		\hat{b}_{\mathbf{k},\sigma}
		\simeq 
		\sum_{m}
		(\mathrm{U} + \mathrm{V})_{\sigma,m}
			\left(
				\hat{a}_{\mathbf{k},m} + \hat{a}_{-\mathbf{k},m}^\dag
			\right)
		\label{negligible}
\end{align}
for $\sigma = \beta$ and $\gamma$.
The corresponding two Bogoliubov operators are then written as 
\begin{widetext}
\begin{align}
	\begin{cases}
		\hat{b}_{\mathbf{k},\beta}
		\simeq
		\mu_{\mathrm{p}} 
			\sum_{m} \zeta_m 
			\left(
				\hat{a}_{\mathbf{k},m} + \hat{a}_{-\mathbf{k},m}^\dag
			\right)
		+
		\mu_{\mathrm{m}} 
			\sum_{m} m \zeta_m 
			\left(
				\hat{a}_{\mathbf{k},m} + \hat{a}_{-\mathbf{k},m}^\dag
			\right),\\
		\hat{b}_{\mathbf{k},\gamma}
		\simeq
		\nu_{\mathrm{p}} 
			\sum_{m} \zeta_m 
			\left(
				\hat{a}_{\mathbf{k},m} + \hat{a}_{-\mathbf{k},m}^\dag
			\right)
		+
		\nu_{\mathrm{m}} 
			\sum_{m} m \zeta_m 
			\left(
				\hat{a}_{\mathbf{k},m} + \hat{a}_{-\mathbf{k},m}^\dag
			\right).
	\end{cases}
	\label{beta_gamma}
\end{align}
\end{widetext}

\begin{figure}[t]
	\begin{center}
	\includegraphics[height=14\baselineskip]{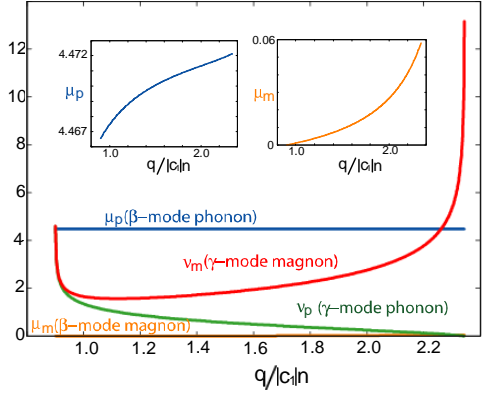}
	\end{center}
	\caption{
(Color online)
Coefficients in the phonon-magnon modes ($\mu_{\mathrm{p}}$,
$\mu_{\mathrm{m}}$, $\nu_{\mathrm{p}}$, and $\nu_{\mathrm{m}}$ in
Eq.~(\ref{beta_gamma})) as functions of the normalized quadratic Zeeman
energy $q / |c_1| n$.
The linear Zeeman energy is chosen to be $ p /|c_1| n = 9 / 10 $.
The vertical axis is scaled by $(\epsilon_{\mathbf{k}}/|c_1| n)^{1/4} /
\sqrt{N_0}$.
The values of the scattering lengths obtained by Kempen \textit{et
al.}~\cite{E.G.M. van Kempen et al.} are used: $a_0=101.8 \,
\mathrm{a.u.}$ and $a_2=100.4 $ a.u.
The insets show the enlarged view of the curves for $\mu_\mathrm{p}$ and
$\mu_\mathrm{m}$.
}
	\label{fig:plot_1.eps}
\end{figure}

Equation (\ref{beta_gamma}) indicates that the quasiparticle operators for
the two gapless modes consist of the number fluctuation operator
	$
		\delta \hat{N} 
		= \sqrt{N_0}
		\left[
			\sum_{m} \zeta_m 
			\left(
				\hat{a}_{\mathbf{k},m} + \hat{a}_{-\mathbf{k},m}^\dag
			\right)
		\right]
	$
and the spin fluctuation operator
	$
		\delta \hat{F}_z 
		= \sqrt{N_0}
		\left[
			\sum_{m} m \zeta_m 
			\left(
				\hat{a}_{\mathbf{k},m} + \hat{a}_{-\mathbf{k},m}^\dag
			\right)
		\right]
	$.
We recall that the operator $\hat{N}$ is the generator of the global phase
rotation and the operator $\hat{F}_z$ is that of the spin rotation around
the $z$ axis.
The creations of the gapless quasiparticles therefore lead to a change in
the global phase of the order parameter and a rotation of the
magnetization around the $z$ axis.
The modes $\beta$ and $\gamma$ are thus the Goldstone modes that restore
the U(1) and SO(2) symmetries.
Since $\delta \hat{N}$ and $\delta \hat{F}_z$ can be regarded as the
phonon and magnon operators, respectively, phonons and magnons are coupled
in the quasiparticles described by Eq.~(\ref{beta_gamma}).
This is in contrast with the cases of the ferromagnetic and polar phases,
in which phonons and magnons are decoupled (see Appendix~\ref{app}).
The numerically obtained coefficients $\mu_{\mathrm{p}}$,
$\mu_{\mathrm{m}}$, $\nu_{\mathrm{p}}$, and $\nu_{\mathrm{m}}$ are shown
in Fig.~\ref{fig:plot_1.eps} as functions of $q / |c_1| n$.
We see that $\hat{b}_{\textbf{k},\beta}$ is mostly the density fluctuation
operator, while $\hat{b}_{\textbf{k},\gamma}$ is the linear combination of
the number and spin fluctuation operators with roughly equal weights for
small $q / |c_1| n$.
In other words, the $\beta$-mode is a phonon-dominant mode and the
$\gamma$-mode is a phonon-magnon coupled mode.
The $\beta$-mode crosses over to a phonon mode across the two neighboring
phase boundaries, while the $\gamma$-mode crosses over to a magnon mode.

\subsection{Coherent excitations}

We investigate the dynamics of the states in which the quasiparticles are
coherently excited.
The excited state is assumed to be a coherent state 
\begin{align}
		| \beta_{k,\sigma} \rangle 
		\equiv 
		e^{
		\beta_{k,\sigma}  \hat{b}_{\mathbf{k},\sigma}^\dag
		-
		\beta_{k,\sigma}^*\hat{b}_{\mathbf{k},\sigma}
		}
		| 0 \rangle_{\mathrm{B}},
\end{align}
where $| 0 \rangle_{\mathrm{B}}$ is the vacuum of the Bogoliubov
quasiparticles.
The change in the expectation value of an observable $\hat{Q}$ due to the
excitation of quasiparticles is given by
\begin{align}
\langle \delta \hat{Q}_{\textbf{k},\sigma}(t) \rangle
=
\langle \beta_{k,\sigma} | \hat{Q}_{\mathrm{H}}(t) | \beta_{k,\sigma} \rangle
- _{\mathrm{B}} \! \langle 0 | \hat{Q}_{\mathrm{H}}(t) | 0
\rangle_{\mathrm{B}},
\label{fluctuation}
\end{align}
where the subscript H denotes the Heisenberg representation.
Since	$\hat{\textbf{A}}_{\textbf{k}} = {}^T \mathrm{U}(k)
\hat{\textbf{B}}_{\textbf{k}} - {}^T \mathrm{V}(k)
\hat{\textbf{B}}_{-\textbf{k}}^*$ from the inverse relation of
Eq.~(\ref{defofqp}), we obtain
\begin{align}
		\langle \delta \hat{\Psi}_{m}(t) \rangle_{\textbf{k},\sigma}
		\nonumber \\
		= \frac{ |\beta_{k,\sigma}| }{ \sqrt{\Omega} } 
		[
			&
			( \mathrm{U} - \mathrm{V} )_{\sigma,m}
			\cos ( \textbf{k} \cdot \textbf{r} - \omega_\sigma t + \phi_{k,\sigma} )
			\nonumber \\
			&+
			i 
			( \mathrm{U} + \mathrm{V} )_{\sigma,m}
			\sin ( \textbf{k} \cdot \textbf{r} - \omega_\sigma t + \phi_{k,\sigma} )
		],
\end{align}
where $\phi_{k,\sigma}=\mathrm{arg}(\beta_{k,\sigma})$ and $\omega_\sigma
= E_{\textbf{k}, \sigma}/\hbar$.
Since the ratio of the real part to the imaginary part is estimated from
Eq.~(\ref{hermitian-like}) to be $(\bar{\epsilon}_{\textbf{k}})^{1/2} \ll
1$, the real part is negligible for the two
gapless modes, $\sigma= \beta$ and $\gamma$, in the long-wavelength
limit.
Therefore, $\langle \delta \hat{\Psi}_{m}(t) \rangle_{\textbf{k},\sigma}$
is almost entirely imaginary, which indicates that the change occurs
mostly in the phase of the real order parameter $\zeta_m$.
Thus, the excitations of $\beta$ and $\gamma$ modes lead to a global phase
rotation and a spin rotation around the $z$ axis.

To study how the quasiparticle excitation rotates the spin, we calculate
$\langle \delta \hat{\textbf{F}}(t) \rangle_{\textbf{k},\sigma}$.
Keeping terms up to those of the first order in $\beta_{k,\sigma}$, we
obtain
\begin{align}
		& \langle \delta \hat{F}_\xi(t) \rangle_{\textbf{k},\sigma}
		\nonumber \\
		& =\frac{ \sqrt{n_0} }{ \sqrt{\Omega} }
		\sum_{m,n} \zeta_m (f_\xi)_{m,n}
		\left[
			( \mathrm{U} \mp \mathrm{V} )_{\sigma,n}
			e^{i( \textbf{k} \cdot \textbf{r} - \omega_\sigma t )}
			\beta_{k, \sigma} \pm \mathrm{H.c.}
		\right],
\end{align}
where $f_\xi$ ($\xi = x, y, z$) are the spin-1 matrices defined in
Eq.~(\ref{fxyz}), and the upper signs refer to $\xi = x$ and $z$ and the
lower signs to $\xi = y$.
Hence, Eq.~(\ref{hermitian-like}) leads to $\langle \delta \hat{F}_y
\rangle_{\textbf{k},\sigma} \gg \langle \delta \hat{F}_x
\rangle_{\textbf{k},\sigma}$ and $\langle \delta \hat{F}_y
\rangle_{\textbf{k},\sigma} \gg \langle \delta \hat{F}_z
\rangle_{\textbf{k},\sigma}$.
This is because $\langle \hat{F}_x \rangle \neq 0$, $\langle \hat{F}_z
\rangle \neq 0$, and $\langle \hat{F}_y \rangle = 0$ from the assumption
of real and positive $\zeta_m$, and the infinitesimal spin rotation around
the $z$ axis changes only $\langle \hat{F}_y \rangle$.
	
\begin{figure}[t]
	\begin{center}
	\includegraphics[height=8\baselineskip]{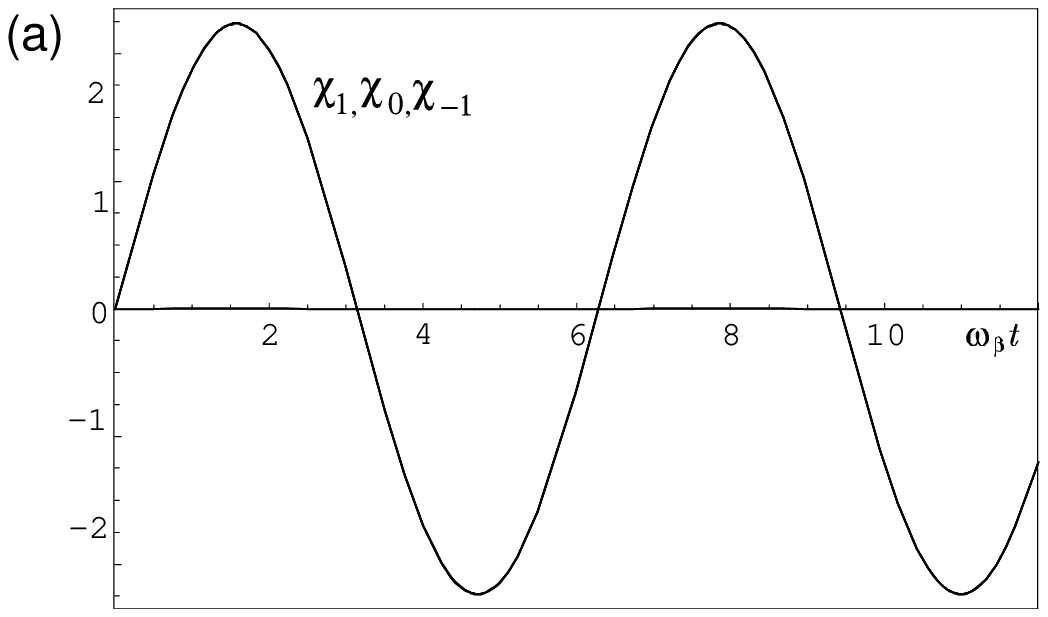}
	\\
	\includegraphics[height=8\baselineskip]{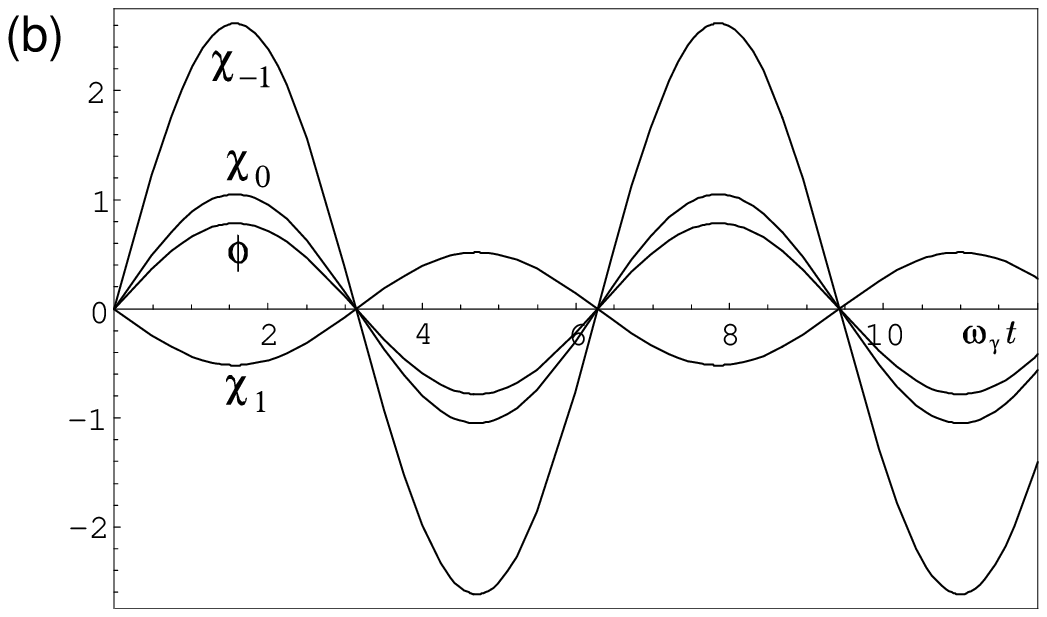}
	\end{center}
	\caption{
Oscillations of the order-parameter phases ($\chi_m \equiv \tan^{-1}
[\mathrm{Im} \langle \hat{\psi}_m \rangle / \mathrm{Re} \langle
\hat{\psi}_m \rangle ]$) and the azimuthal angle of magnetization around
the $z$ axis ($\phi \equiv \tan^{-1} [\langle \hat{F}_y \rangle / \langle
\hat{F}_x \rangle ]$) caused by the excitation of the Bogoliubov
quasiparticles in the long-wavelength limit.
The vertical axis is marked in arbitrary units.
The Zeeman energies are assumed to be $p/|c_1| n = 9/10$ and $q/|c_1| n =
11/10$.
(a) Phonon-dominant mode ($\beta$-mode). 
Since rotations of the order-parameter phases are much larger than that of
the magnetization, we cannot see the latter in this figure.
(b) Phonon-magnon coupled mode ($\gamma$-mode).
}
\label{fig:angle fluc}
\end{figure}

Thus, we have shown that the excitations of the Goldstone modes $\beta$
and $\gamma$ lead to U(1) and SO(2) transformations.
Oscillations of the order-parameter phases and those of the azimuthal
angle of the spin vector are shown in Fig.~\ref{fig:angle fluc}.
Figure~\ref{fig:angle fluc}(a) shows that the $\beta$-mode excitation
changes the phases of $\langle \hat\psi_1 \rangle$, $\langle \hat\psi_0
\rangle$, and $\langle \hat\psi_{-1} \rangle$ in the same manner.
This is because as shown in Fig.~\ref{fig:plot_1.eps} the dominant
contribution to the $\beta$ mode is made by phonons which are insensitive
to individual spin components.
On the other hand, the $\gamma$-mode excitation describes not only
fluctuations of the overall phase but also those of the spin vector around
the $z$ axis.
Since the rotation around the $z$ axis is $e^{i \varphi \hat{F}_z}| m
\rangle = e^{i \varphi m}| m \rangle$, $\chi_1$ and $\chi_{-1}$ are out of
phase with respect to $\chi_0$.
 
\begin{figure}[t]
	\begin{center}
	\includegraphics[height=10\baselineskip,width=15\baselineskip]{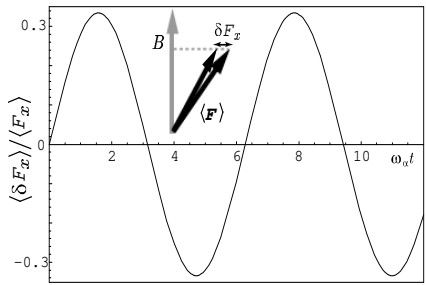}
	\end{center}
	\caption{
Oscillation of transverse magnetization $\langle F_x \rangle$ due to
coherent excitation of the gapful mode ($\alpha$-mode).
The inset schematically illustrates the change in the spin vector caused
by the excitation of the $\alpha$-mode.
}
\label{fig:wavealpha.eps}
\end{figure}

The gapful mode ($\alpha$-mode) can be interpreted as playing the role of
changing the magnitude of magnetization.
As shown in Fig.~\ref{fig:wavealpha.eps}, the fluctuation of $F_x$ is
dominant when the $\alpha$-mode is excited.
The $z$ component $F_z$ cannot vary due to the spin conservation, and
hence the spin fluctuation is restricted in the $x$-$y$ plane as
illustrated in the inset of Fig.~\ref{fig:wavealpha.eps}.
			
\section{Discussion} \label{sectiond}

We have shown that the ground state in the shaded region of
Fig.~\ref{fig:eps.eps} is the broken-axisymmetry phase featuring
transverse magnetization.
Here, we discuss possible experimental consequences of the axisymmetry
breaking and the transverse magnetization.

Let us consider a situation in which atoms are prepared in the $m = 0$
state.
When $p = 0$ and $q < 2 |c_1| n$, the ground state is in the
broken-axisymmetry phase and the transverse magnetization should develop
in time.
However, the total spin angular momentum parallel to the magnetic field
must be conserved, and for small $q$ the magnitude of the transverse
component of the total spin is nearly entirely conserved.
Consequently, local magnetization varies in space only insofar as the
total spin is conserved~\cite{Saito,SaitoL}.
This constraint leads to formation of various spin textures depending on
the trap geometry.
For example, in an elongated cigar-shaped trap, a staggered domain
structure or a helical structure is formed spontaneously~\cite{Saito}.
In a pancake-shaped trap, on the other hand, a concentric domain structure
is formed~\cite{Saito}.

The formation of the domain structure costs the kinetic energy and the
ferromagnetic energy at the domain walls.
If the direction of the spin vector changes gradually over space, the
formation of the spin texture costs little energy.
One of such textures is a topological spin texture, in which the
orientation of transverse magnetization has a $2\pi$ winding about a
central defect.
When the size of the system is small, the topological spin texture becomes
energetically favorable and develops spontaneously~\cite{SaitoL}.
Recently, formation of the topological spin texture has been observed by
the Berkeley group~\cite{Sadler}, in which the state of the system is
changed rapidly by a change in the magnetic field from the $| 0 \rangle$
region to the shaded region in Fig.~\ref{fig:eps.eps}.
The spontaneous transverse magnetization in the Berkeley experiment is a
manifestation of the symmetry breaking discussed in the present paper.
If the amount of the change in the magnetic field is small or if the speed
of the change is slow, the system is not markedly disturbed, and
low-energy gapless excitations (the $\beta$ and $\gamma$ modes) should be
observed.

\section{Conclusions} \label{section4}

We have studied a spin-1 ferromagnetic BEC by taking the quadratic Zeeman
effect into account.
The mean field theory predicts that BECs with ferromagnetic interactions 
show the broken-axisymmetry phase, in which magnetization tilts against
the direction of an external magnetic field.
Here, the SO(2) symmetry is broken, in addition to the U(1) global phase
symmetry.
Applying the Bogoliubov theory for a BEC with spin degrees of freedom, we
have found one gapful mode and two gapless Goldstone modes for this
phase.
We have analytically shown that two gapless modes are the coupled
phonon-magnon modes that restore the U(1) and SO(2) symmetries
simultaneously.
Numerical analysis has shown that one Goldstone mode is the
phonon-dominant mode and the other is the phonon-magnon coupled mode with
roughly equal weights.
The gapful mode changes the length of the spin by fluctuating the spin in
the direction perpendicular to the magnetic field (see the inset of
Fig.~\ref{fig:wavealpha.eps}).

When more than one continuous symmetry is spontaneously broken, multiple
gapless modes emerge and they couple with each other to form Goldstone
modes as shown in this paper.
Such multiple Goldstone modes may therefore be found in spin-2
BECs~\cite{Ueda&Koashi} and higher spin BECs, which merit further
investigation.

\section*{Acknowledgements}

We thank T.~Mori for his participation in an early stage of this work.
This work was supported by Grants-in-Aid for Scientific Research 
(Grant No.~17071005 and No.~17740263)
and by the 21st Century COE programs on ``Nanometer-Scale Quantum
Physics'' and ``Coherent Optical Science'' from the Ministry of Education,
Culture, Sports, Science and Technology of Japan.
M.U. acknowledges support by a CREST program of the JST.

\appendix
\section{Derivation of the phase diagram for $c_1 < 0$}
\label{app0}

In this appendix, we derive the analytic expression for the spin
components in Eq.~(\ref{mixed}) in the broken-axisymmetry phase and
reproduce the phase diagram for $c_1 < 0$ as shown in
Fig.~\ref{fig:eps.eps}.

The average energy per atom is given from Eq.~(\ref{Hbec}) by,
\begin{eqnarray}
\label{h}
e & \equiv & \frac{\left\langle \hat{\mathcal{H}}_{\mathrm{BEC}}
\right\rangle}{N_0} \nonumber \\
& = & \frac{|c_1| n}{2} \left| 2\zeta_1 \zeta_{-1} -
\zeta_0^2 \right|^2 + \sum_{m = -1}^{1} (-p m + q m^2) |\zeta_m|^2.
\nonumber \\
\end{eqnarray}
The first term on the right-hand side of Eq.~(\ref{h}) can be rewritten as
\begin{eqnarray}
& & \frac{|c_1| n}{2} \left| 2\zeta_1 \zeta_{-1} - \zeta_0^2 \right|^2 
\nonumber \\
& = &
\frac{|c_1| n}{2} \left| 2|\zeta_1 \zeta_{-1}| - |\zeta_0|^2 e^{i (2\chi_0
- \chi_1 - \chi_{-1})} \right|^2,
\end{eqnarray}
where $\zeta_m = |\zeta_m| e^{i \chi_m}$.
Hence, the phase depends on energy only through $2\chi_0 - \chi_1 -
\chi_{-1}$, and the energy is minimized for $2\chi_0 - \chi_1 - \chi_{-1}
= 0$.
Without loss of generality, we assume that $\zeta_m$'s are real and
non-negative.

We minimize
\begin{equation}
K \equiv e - \mu (\zeta_1^2 + \zeta_0^2 + \zeta_{-1}^2),
\end{equation}
where a Lagrange multiplier $\mu$ is introduced to ensure the
normalization condition $\zeta_1^2 + \zeta_0^2 + \zeta_{-1}^2 = 1$.
Stationary conditions can be obtained through differentiation of $K$ with
respect to $\zeta_m$'s as
\begin{eqnarray}
\frac{\partial K}{\partial \zeta_{\pm 1}} & = & 2 |c_1| n (2 \zeta_1
\zeta_{-1} - \zeta_0^2) \zeta_{\mp 1} + 2 (q \mp p - \mu) \zeta_{\pm
1} = 0, \nonumber \\
\label{Kz1}
\\
\frac{\partial K}{\partial \zeta_0} & = & -2 |c_1| n (2 \zeta_1 \zeta_{-1}
- \zeta_0^2) \zeta_0 - 2 \mu \zeta_0 = 0.
\label{Kz0}
\end{eqnarray}
It follows from Eq.~(\ref{Kz0}) that either $\zeta_0 = 0$ or
\begin{equation} \label{mu}
\mu = -|c_1| n (2 \zeta_1 \zeta_{-1} - \zeta_0^2)
\end{equation}
has to be satisfied.
When $\zeta_0 = 0$, we find that $K$ is minimized with
\begin{eqnarray}
\zeta_1 = 1, \; \zeta_{-1} = 0 \;\; \mbox{for $p > 0$} \;\;\;\;
(h = q - p), \\
\zeta_1 = 0, \; \zeta_{-1} = 1 \;\; \mbox{for $p < 0$} \;\;\;\;
(h = q + p).
\end{eqnarray}
When $\zeta_0 \neq 0$, Eq.~(\ref{Kz1}) becomes
\begin{equation} \label{lz1}
(q \mp p - \mu) \zeta_{\pm 1} = \mu \zeta_{\mp 1}.
\end{equation}
We can easily see that the polar state
\begin{equation}
\zeta_0 = 1, \; \zeta_1 = \zeta_{-1} = 0 \;\;\;\; (h = 1)
\end{equation}
satisfies Eq.~(\ref{lz1}).
Using Eqs.~(\ref{mu}) and (\ref{lz1}), we obtain the
solution for the broken-axisymmetry phase in Eq.~(\ref{mixed}).
This solution is valid for
\begin{equation} \label{region}
q^2 - p^2 \geq 0 \;\; \mbox{and} \;\; 2 |c_1| n q - q^2 + p^2 \geq 0.
\end{equation}
The energy of the broken-axisymmetry state is calculated to be
\begin{equation}
e_{\rm br} = \frac{1}{4 (|c_1| n q)^2} (q^2 - p^2)(p^2 - 4 |c_1| n q -
q^2).
\end{equation}
We can show that $e_{\rm br} \leq 1$ and $e_{\rm br} \leq q \pm p$ are
always satisfied in the region of Eq.~(\ref{region}), and thus we obtain
the phase diagram in Fig.~\ref{fig:eps.eps}.

The three components of the spin-1 matrices are given by
\begin{eqnarray} \label{fxyz}
f_x & = & \frac{1}{\sqrt{2}} \left( \begin{array}{ccc} 0 & 1 & 0 \\ 1 & 0 & 1
\\ 0 & 1 & 0 \end{array} \right),
f_y = \frac{i}{\sqrt{2}} \left( \begin{array}{ccc} 0 & -1 & 0 \\ 1 & 0 &
-1 \\ 0 & 1 & 0 \end{array} \right),
\nonumber \\
f_z & = & \left( \begin{array}{ccc} 1 & 0 & 0 \\ 0 & 0 & 0 \\ 0 & 0 & -1
\end{array} \right),
\end{eqnarray}
and the corresponding spin components are given by
\begin{eqnarray}
\left\langle F_x \right\rangle & \equiv & N_0 \bm{\zeta}^\dagger f_x
\bm{\zeta} = \frac{N_0}{\sqrt{2}} \left[ \zeta_0^* \left( \zeta_1 +
\zeta_{-1} \right) + \zeta_0 \left( \zeta_1^* + \zeta_{-1}^* \right)
\right], \nonumber \\
\label{fxdef} \\
\left\langle F_y \right\rangle & \equiv & N_0 \bm{\zeta}^\dagger f_y
\bm{\zeta} = i \frac{N_0}{\sqrt{2}} \left[ \zeta_0^* \left( \zeta_1 -
\zeta_{-1} \right) - \zeta_0 \left( \zeta_1^* - \zeta_{-1}^* \right)
\right], \nonumber \\
\\
\left\langle F_z \right\rangle & \equiv & N_0 \bm{\zeta}^\dagger f_z
\bm{\zeta} = N_0 \left( |\zeta_1|^2 - |\zeta_{-1}|^2 \right).
\label{fzdef}
\end{eqnarray}
Substituting Eq.~(\ref{mixed}) into Eqs.~(\ref{fxdef})-(\ref{fzdef}), we
obtain Eqs.~(\ref{fperp})-(\ref{fmag}).

\section{Excitation spectra in the ferromagnetic and polar phases}
\label{app}

\subsection{Ferromagnetic phase}
We first consider the ferromagnetic phase ($|+1\rangle$ in
Fig.~\ref{fig:eps.eps}) for $p>0$.
In this phase the order parameters are given by $\zeta_1 = 1$ and $\zeta_0
= \zeta_{-1} = 0$.
The matrices M and N in Eq.~(\ref{differential of a}) are shown to be
\begin{eqnarray}
\mathrm{M} & = & \begin{pmatrix}
\epsilon_{\mathbf{k}} + c_0 n + c_1 n & 0 & 0 \\
0 & \epsilon_{\mathbf{k}} + p - q & 0 \\
0 & 0 & \epsilon_{\mathbf{k}} + 2p - 2c_1 n
\end{pmatrix}, \nonumber \\
\\
\mathrm{N} & = & \begin{pmatrix}
c_0 n + c_1 n & 0 & 0 \\
0 & 0 & 0 \\
0 & 0 & 0
\end{pmatrix},
\end{eqnarray}
and the matrix G in Eq.~(\ref{defG}) is given by
\begin{align}
		\mathrm{G} = 
		\begin{pmatrix}
			\epsilon_{\mathbf{k}}( \epsilon_{\mathbf{k}} + 2 g_2 n) & 0 & 0 \\
			0 & (\epsilon_{\mathbf{k}} + p - q )^2 & 0 \\
			0 & 0 & ( \epsilon_{\mathbf{k}} - 2p - 2 c_1 n )^2
		\end{pmatrix}.
\end{align}
Diagonalizing G we obtain three excitation energies as
\begin{align}
		\begin{cases}
			E_{\mathrm{p}} = \sqrt{
				\epsilon_{\mathbf{k}}( \epsilon_{\mathbf{k}} + 2 g_2 n)
				},\\
			E_{0} = \epsilon_{\mathbf{k}} + p - q,\\
			E_{-1} = \epsilon_{\mathbf{k}} - 2p - 2 c_1 n,
	    \end{cases}
\label{Eferro}
\end{align}
and the associated quasiparticle operators as
\begin{align}
\begin{cases}
		\hat{b}_{\mathbf{k}, \mathrm{p}}
		= \frac
		{
			\sqrt{
			\epsilon_{\mathbf{k}} + g_2 n + E_{\mathrm{p}}
			} 
		}
		{\sqrt{2 E_{\mathrm{p}}} }
		\hat{a}_{\mathbf{k}, 1}
		+
		\frac
		{
			\sqrt{
			\epsilon_{\mathbf{k}} + g_2 n - E_{\mathrm{p}}
			} 
		}
		{\sqrt{2 E_{\mathrm{p}}} }
		\hat{a}_{-\mathbf{k}, 1}^\dagger, \\
\hat{b}_{\mathbf{k}, 0} = \hat{a}_{\mathbf{k}, 0}, \\
\hat{b}_{\mathbf{k}, -1} = \hat{a}_{\mathbf{k}, -1}.
\end{cases}
\label{phonon exc.}
\end{align}
The corresponding results for $p<0$ can be obtained by changing the linear
Zeeman energy $p$ to $-p$.
For $q = 0$ these results reduce to those obtained in the absence of the
quadratic Zeeman effect~\cite{M.Ueda}.
The excitation spectrum $E_{\mathrm{p}}$ is similar to the one obtained
for scalar BECs;
it is a phonon mode with the speed of sound given by $c = \sqrt{g_2 n /
M}$.
The other two modes are understood as excitations from the $m = 1$ state
to the $m = 0$ and $m = -1$ states, respectively.
In the long-wavelength limit ($k \to 0$), $E_0$ ($E_{-1}$) coincides with
the single-particle energy difference between the $m = 1$ state and the $m
= 0$ ($m = -1$) state.
Hence, the phonon mode ($E_p$) and the magnon modes ($E_0, E_{-1}$) are
decoupled in the ferromagnetic phase.

It can be easily checked that these excitation modes and operators do
reproduce the effective Hamiltonian (\ref{original Hamiltonian}) through
Eq.~(\ref{free particle like}).
For small $\epsilon_\textbf{k}$, $E_{\mathrm{p}}$ in Eq.~(\ref{Eferro})
is proportional to $\epsilon_\textbf{k}^{1/2}$ and $\hat{b}_{\mathbf{k},
\mathrm{p}}$ in Eq.~(\ref{phonon exc.}) is proportional to
$\epsilon_{\textbf{k}}^{-1/4}$.
The singular $\epsilon_{\textbf{k}}$-dependence in each of
$E_{\mathrm{p}}$ and $\hat{b}_{\mathbf{k}, \mathrm{p}}$ is therefore
canceled in the product $E_{\mathrm{p}} \hat{b}^\dagger_{\mathbf{k},
\mathrm{p}} \hat{b}_{\mathbf{k}, \mathrm{p}}$, giving the original
$\epsilon_{\textbf{k}}$-dependence in the effective Hamiltonian.

\subsection{Polar phase}

In the polar phase ($|0\rangle$ in Fig.~\ref{fig:eps.eps}), the order
parameters are $\zeta_0 = 1$ and $\zeta_1 = \zeta_{-1} = 0$, and the
matrices M and N have the forms
\begin{eqnarray}
\mathrm{M} & = & \begin{pmatrix}
\epsilon_{ \textbf{k} } - p + q + c_1 n & 0 & \\
0 & \epsilon_{ \textbf{k} } + c_0 n & 0 \\
0 & 0 & \epsilon_{ \textbf{k} } + p + q + c_1 n
\end{pmatrix}, \nonumber \\
\\
\mathrm{N} & = & \begin{pmatrix}
0 & 0 & c_1 n \\
0 & c_0 n & 0 \\
c_1 n & 0 & 0
\end{pmatrix},
\end{eqnarray}
which give
\begin{widetext}
\begin{align}
		\mathrm{G} = 	
			\begin{pmatrix}
			(\epsilon_{ \textbf{k} } - p + q + c_1 n)^2 - c_1 n^2 & 0 & 2 c_1 n p \\
			0 & \epsilon_{ \textbf{k} } (\epsilon_{ \textbf{k} } + 2 c_0 n) & 0 \\
			-2 c_1 n p & 0 & (\epsilon_{ \textbf{k} } + p + q + c_1 n)^2 - c_1 n^2
			\end{pmatrix}.
\end{align}
The eigenvalues are obtained as 
\begin{align}
			\begin{cases}
				E_{\mathrm{p}} = \sqrt{
					\epsilon_{\textbf{k}}( \epsilon_{\textbf{k}} + 2 c_0 n)
					},\\
				E_{ + } = -p 
					+ \xi_\textbf{k}, \\
				E_{ - } = +p 
					+ \xi_\textbf{k},
		    \end{cases}
		   \label{Exc. polar}
	\end{align}
where 
$
		\xi_\textbf{k} = 
		\sqrt{
				(\epsilon_{\textbf{k}} + q)
				(\epsilon_{\textbf{k}} + q + 2 c_1 n) 
			}
$.
The quasiparticle operators are found to be
\begin{align}
	\begin{cases}
		\displaystyle
		\hat{b}_{\mathbf{k}, \mathrm{p}}
		= \frac
		{
			\sqrt{
			\epsilon_{\mathbf{k}} + c_0 n + E_{\mathrm{p}}
			} 
		}
		{\sqrt{2 E_{\mathrm{p}}} }
		\hat{a}_{\mathbf{k}, 0}
		+
		\frac
		{
			\sqrt{
			\epsilon_{\mathbf{k}} + c_0 n - E_{\mathrm{p}}
			} 
		}
		{\sqrt{2 E_{\mathrm{p}}} }
			\hat{a}_{-\mathbf{k}, 0}^\dag ,\\
		\displaystyle
		\hat{b}_{\mathbf{k}, \mathrm{\pm}}
		= 
		\frac
		{
			\sqrt{
				\epsilon_{\textbf{k}} + q + c_1 n + \xi_{\textbf{k}}
			} 
		}
		{\sqrt{2 \xi_{\textbf{k}} } }
		\hat{a}_{\mathbf{k}, \pm1}
		-
		\frac
		{
			\sqrt{
				\epsilon_{\textbf{k}} + q + c_1 n - \xi_{\textbf{k}}
			} 
		}
		{\sqrt{2 \xi_{\textbf{k}} } }
		\hat{a}_{-\mathbf{k}, \mp1}^\dag.
	\end{cases}
	\label{Opr. polar}
\end{align}
\end{widetext}	
The first mode $E_{\mathrm{p}}$ in Eq.~(\ref{Exc. polar}) is a phonon
mode.
The other two are magnon modes which scatter a quasiparticle from the
state with $m=0$ to the one with $m = \pm 1$.
Thus, also in this phase, phonons and magnons are decoupled.


\begin{thebibliography}{99}

\bibitem{D.M.Stamper-Kurn et al.} D.M. Stamper-Kurn, M. R. Andrews, A.P. Chikkatur,
	S. Inouye, H.-J. Miesner, J. Stenger, and W. Ketterle,
	Phys. Rev. Lett. \textbf{80}, 2027 (1998). 
\bibitem{J.Stenger et al.} J. Stenger, S. Inouye, D.M. Stamper-Kurn, H.-J. Miesner,
	A.P. Chikkatur, and W. Ketterle,
	Nature (London) \textbf{396}, 345 (1998).
\bibitem{H.-J.Miesnr et al.} H.-J. Miesner, D. M. Stamper-Kurn, J. Stenger, S. Inouye,
	A. P. Chikkatur, and W. Ketterle,
	Phys. Rev. Lett. \textbf{82}, 2228 (1999).
\bibitem{tunneling} D. M. Stamper-Kurn, H.-J. Miesner, A. P. Chikkatur, S. Inouye,
	J. Stenger, and W. Ketterle,
	Phys. Rev. Lett. \textbf{83} 661 (1999).
\bibitem{A.Gorllitz et al.} A. G\"{o}rlitz, T. L. Gustavson, A. E. Leanhardt, 
	R. L\"{o}w, A. P. Chikkatur, S. Gupta, S. Inouye, D. E. Pritchard, and W. Ketterle,
	Phys. Rev. Lett. \textbf{90}, 090401 (2003). 
\bibitem{H.Schmaljohann et al.} H. Schmaljohann, M. Erhard, J. Kr\"{o}njager,
	M. Kottke, S. van Staa, L. Cacciapuoti, J. J. Arlt, K. Bongs, and K. Sengstock,
	Phys. Rev. Lett. \textbf{92}, 040402 (2004). 
\bibitem{M.-S. Chang et al.} M.-S. Chang, C. D. Hamley, M. D. Barrett, J. A. Sauer, K. M. Fortier,
	W. Zhang, L. You, and M. S. Chapman, 
	Phys. Rev. Lett. \textbf{92}, 140403 (2004). 
\bibitem{T.Kuwamoto et al.} T. Kuwamoto, K. Araki, T. Eno, and T. Hirano, 
	Phys. Rev. A \textbf{69}, 063604 (2004). 
\bibitem{T-L Ho} T.-L. Ho, 
	Phys. Rev. Lett. \textbf{81}, 742 (1998).
\bibitem{T.Ohmi&K.Machida} T. Ohmi and K. Machida, 
	J. Phys. Soc. Jpn \textbf{67}, 1822 (1998). 
\bibitem{Ciobanu et al.} C. V. Ciobanu, S. K. Yip, and T.-L. Ho,
	Phys. Rev. A \textbf{61}, 033607 (2000).
\bibitem{Ueda&Koashi} M. Ueda and M. Koashi,
	Phys. Rev. A \textbf{65}, 063602 (2002).
\bibitem{Law et al.} C. K. Law, H. Pu, and N. P. Bigelow, 
	Phys. Rev. Lett. \textbf{81}, 5257 (1998).
\bibitem{M.Koashi&M.Ueda} M. Koashi and M. Ueda,
	Phys. Rev. Lett. \textbf{84}, 1066 (2000).
\bibitem{Ho&Yip} T.-L. Ho and S. K. Yip, 
	Phys. Rev. Lett. \textbf{84}, 4031 (2000).
\bibitem{Huang&Gou} W.-J. Huang and S.-C. Gou,
	Phys. Rev. A \textbf{59}, 4608 (1999) .
\bibitem{M.Ueda} M. Ueda, 
	Phys. Rev. A \textbf{63}, 013601 (2000).
\bibitem{Peter et al.} P. Sz\'{e}pfalusy and G. Szirmai,	
	Phys. Rev. A \textbf{65} 043602 (2002).
\bibitem{Klausen et al.} N. N. Klausen, J. L. Bohn and C. H. Greene,
	Phys. Rev. A \textbf{64}, 053602 (2001).
\bibitem{E.G.M. van Kempen et al.} E. G. M. van Kempen, S. J. J. M. F. Kokkelmans, D. J. Heinzen, 
	and B. J. Verhaar, 
	Phys. Rev. Lett. \textbf{88}, 093201 (2002). 
\bibitem{Goldstone} J. Goldstone, 
	Nuovo Cimento \textbf{19}, 154 (1961).
\bibitem{Bogoliubov}
	N. N. Bogoliubov, 
	J. Phys. USSR \textbf{11}, 23 (1947).
\bibitem{Saito}
H. Saito and M. Ueda, Phys. Rev. A {\bf 72}, 023610 (2005).
\bibitem{SaitoL}
H. Saito, Y. Kawaguchi, and M. Ueda, Phys. Rev. Lett. {\bf 96}, 065302
(2006).
\bibitem{Sadler}
L. E. Sadler, J. M. Higbie, S. R. Leslie, M. Vengalattore, and
D. M. Stamper-Kurn, cond-mat/0605351.

\end{thebibliography}
\end{document}